# An Enhanced Middleware for Collaborative Privacy in IPTV Recommender Services

Ahmed M. Elmisery* and Dmitri Botvich
Telecommunications Software & Systems Group-TSSG
Waterford Institute of Technology-WIT, Co. Waterford, Ireland
*ahmedmohmed2001@gmail.com

*Abstract*—One of the concerns users have to confronted when using IPTV system is information overload that makes it difficult for them to find a suitable content according to their personal preferences. Recommendation service is one of the most widely adopted technologies to alleviating this problem; these services intend to provide people with referrals of items they will appreciate based upon their preferences. IPTV users must ensure their sensitive preferences collected by any recommendation service are properly secured. In this work, we introduce a framework for private recommender service based on Enhanced Middleware for Collaborative Privacy (*EMCP*). *EMCP* executes a two-stage concealment process that gives the user a complete control on the privacy level of his profile. We utilize trust mechanism to augment recommendation's accuracy and privacy. Trust heuristic spot users whom are trustworthy with respect to the user requesting recommendation (target-user). Later, the neighbourhood formation is calculated using proximity metrics based on these trustworthy users. Finally, Users submit their profiles in an obfuscated form without revealing any information about their data, and the computation of recommendations proceeds over the obfuscated data using secure multi-party computation protocol. We expand the obfuscation scope from single obfuscation level for all users to arbitrary obfuscation levels based on trustworthy between users. In other words, we correlate the obfuscation level with different trust levels, so the more trusted a target user is the less obfuscation copy of users' profile he can access. We also provide an IPTV network scenario and experimentation results. Our results and analysis shows that our two-stage concealment process not only protect the users' privacy, but also can maintain the recommendations accuracy.

## I. INTRODUCTION

Internet Protocol Television (IPTV) is a video service providing IP broadcasts and video on demand (VOD) over a broadband IP content delivery network (CDN) specialized in video services. The IPTV user has access to myriads of video content spanning IP Broadcast and VOD [1]. In this context, it is difficult for them to find content that matches their preferences from the huge amount of video content available. In order to attract and satisfy these users, IPTV service providers employ recommendation services to increase their revenues and offer added value to their patrons. Recommendation service is a promising personalized service for IPTV service which offer referrals to users by building up users' profiles (explicit or implicit) based on their past ratings, behaviour, purchase history or demographic information. In the context of this work, a profile is a list comprises the video contents the user has watched or purchased combined with their meta-data extracted from the content provider (i.e. genres, directors, actors and so on) and the ratings the user gave to these contents.

Recommendation services are usually served using collaborative filtering (CF) algorithms, which is a popularly used technical approach to automate the word-of-mouth process; it is based on the hypothesis that people with similar tastes prefer the same items. The recommendations using CF technique involves a server or main entity that collects users' profiles to find users similar to the user receiving the recommendation (target user) and then it executes CF algorithms to suggest to him/ her items rated high in the past by them. Because data collected from IPTV users cover personal information about different contents they watched or purchased, there is a serious threat to individual privacy. This data can be used for unsolicited marketing, government surveillance, profiling users, misused or it can be sold by providers when they suffered bankruptcy.

While In the general case, collecting high quality profiles from users is desirable as such the recommendations can be highly beneficial both for the users and the IPTV service providers, but it is not an easy task as the price is likewise high: total loss of privacy while generating recommendations. On one hand, some users are willing to reveal their whole profiles in order to get accurate referrals but others may be concerned about the privacy implications of disclosing their profiles which can open a door for the misuse of personal data. Currently there are two options for privacy concerned users when using IPTV recommender service: first, they can refuse to enter the information they are uncomfortable about disclosing which brings the sparse data problem [2] for the recommendation technique, since only a subset of items have ratings scored by the user. Second, they enter false information which decreases the accuracy of the generated recommendations, this result to the lack of acceptance of the respective services in general. As a matter of fact, an actual rating given to an item by user produces a reasonable explanation and rank from a reliable source. Users are more likely willing to give more truthful data if privacy measurements are provided or if they assured that the data does not leave their personal devices until it is properly desensitised.

In this work, we present an enhanced middleware for collaborative privacy (*EMCP*) that allows creating serendipity recommendations without breaching users' privacy. *EMCP* employs a set of mechanisms to allow users to share their data among each other in the network to attain collaborative privacy. The users' cooperation is needed not only to protect their privacy but also to make the service run properly. This approach preserves the aggregates in the dataset to maximize the usability of information in order to accurately predicate ratings for items that have not consumed before by the target-user. Two novel mechanisms used in *EMCP* to secure the user rating profile in the untrusted PRS with minimum loss of accuracy. The first mechanism based on an algorithm called Clustering Transformation Algorithm (*CTA*) for obfuscating the user rating profile before sharing it with other users in the IPTV network. It partitions the user rating profile into smaller clusters and then obfuscates each cluster in a way to preserve the distances between data points inside the same cluster. The second mechanism based on a protocol called Secure Recommendation (*SR*) that is build upon Paillier scheme of homomorphic encryption in order to permit particular operations to be performed on encrypted data without need for prior decryption. This means that, we can retrieve the original statistical properties without using raw user's data. In addition, *EMCP* employs interpersonal trust between users to enhance recommendations' accuracy and preserve privacy. The enhancement in accuracy is achieved by employing trust based heuristics to propagate and spot users whom are trustworthy with respect to the target user. Moreover, trust based heuristics enhance privacy by transforming participants' data in different ways based on different trust levels to hide the raw ratings. Thus, In contrast to a single obfuscation level scenario where only one obfuscated copy is released for all users using fixed parameters for the obfuscation mechanism, now multiple differently obfuscated copies of the same data is released for different users with different trust levels. The more trusted the user is the less obfuscated copy he can access. These different copies can be generated in various fashions. They can be jointly generated at different times upon receiving new request from target user, or on demand fashion. The latter case gives users maximum flexibility.

In rest of this work, we will generically refer to news programs, movies and video on demand contents as Items. This paper is organised as follows. In Section II, related works are described. Section III introduces IPTV network scenario landing our private recommender service. The proposed solution based on *EMCP* is introduced in Section IV. In Section V, The two-stage concealment process is described in details. Proof of security and correctness for the two-stage concealment process is demonstrated in Section VI. In Section VII, the Results from some experiments on the proposed mechanisms are reported. Finally, the conclusions and future work are given in Section VIII.

## II. RELATED WORKS

Majority of the existing Recommender services are based on collaborative filtering, others focus on content based filtering using EPG data [3]. Collaborative filtering techniques build users' profiles in two ways upon ratings (explicit ratings procedures) or log archives (implicit ratings procedures) [4]. These procedures lead to two different approaches for the collaborative filtering including the rating based approaches and log based approaches. The majority of the literature addresses the problem of privacy on collaborative filtering technique, due to it is a potential source of leakage of private information shared by the users as shown in [5]. In [6] It is proposed a theoretical framework to preserve privacy of customers and the commercial interests of merchants. Their system is a hybrid recommender system that uses secure two party protocols and public key infrastructure to achieve the desired goals. In [7, 8] it is proposed a privacy preserving approach based on peer to peer techniques using users' communities, where the community will have a aggregate user profile representing the group as whole but not individual users. Personal information will be encrypted and the communication will be between individual users but not servers. Thus, the recommendations will be generated at client side. In [9, 10] it is suggested another method for privacy preserving on centralized recommender systems by adding uncertainty to the data by using a randomized perturbation technique while attempting to make sure that necessary statistical aggregates such as the mean don't get disturbed much. Hence, the server has no knowledge about true values of individual rating profiles for each user. They demonstrate that this method does not decrease essentially the obtained accuracy of the results. But recent research work [11, 12] pointed out that these techniques don't provide levels of privacy as it was previously thought. In [12] it is pointed out that arbitrary randomization is not safe because it is easy to breach the privacy protection it offers. They proposed a random matrix based spectral filtering techniques to recover the original data from perturbed data. Their experiments revealed that in many cases random perturbation techniques preserve very little privacy. Similar limitations were detailed in [11]. Storing user's rating profiles on their own side and running the recommender system in distributed manner without relying on any server is another approach proposed in [13], where authors proposed transmitting only similarity measures over the network and keep users rating profiles secret on their side to preserve privacy. Although this method eliminates the main source of threat against user's privacy, but it requires higher cooperation among users to generate useful recommendations. The work in [14] stated that existing similarities deem useless as traditional user profiles are sparse and insufficient. Recommender systems need new ways to calculate user similarities. They utilize interpersonal trustworthiness to describe the relationship between two users. The authors in [15] shows the correlation between similarity and trust and how it can elevate movie recommendations accuracy.

In this work, *EMCP* preserves the privacy of user rating profile form the attack model presented in [16]. The attack model for data obfuscation is different from the attack model for encryption-based techniques, but no common standard has been implemented for data obfuscation. Existing attack models has primarily considered a case where the attacker correlates obfuscated data with data from other publicly-accessible databases in order to reveal the sensitive information. But the attack model presented in [16], considers a case where the attacker colludes with some users in the network to obtain some partial information about the process used to obfuscate the data and/or some of the original data items themselves. The attacker can then use this partial information to attempt to reverse engineer the entire dataset.

## III. PRIVATE RECOMMENDER SERVICE FOR IPTV NETWORK SCENIRO

We extend the scenario proposed in [17-21], where a private recommender service (PRS) is implemented as an external third party server and users gave their rating profiles to that server in order to receive recommendations. The basic idea for recommendation based on *EMCP* is as follows: Upon receiving a request from target user, a group of participants is formed that is managed by an elected super-peer. Each participant obfuscates his ratings profile using a multi-level obfuscation mechanism provided by *EMCP*, such that each profile is obfuscated based on the estimated trust level with the target user, further more this step prevents the super-peers from learning each participant's raw ratings. The super-peer collects these obfuscated rating profiles and computes an aggregation on it, which does not expose individual ratings. Next, the aggregated data is encapsulated using doubly encryption and submitted to PRS to predicate ratings for recommended items that will be offered in the end to the target-user. The collaborative filtering task at PRS will be reduced to computing addition on aggregated data without exposing the raw data.

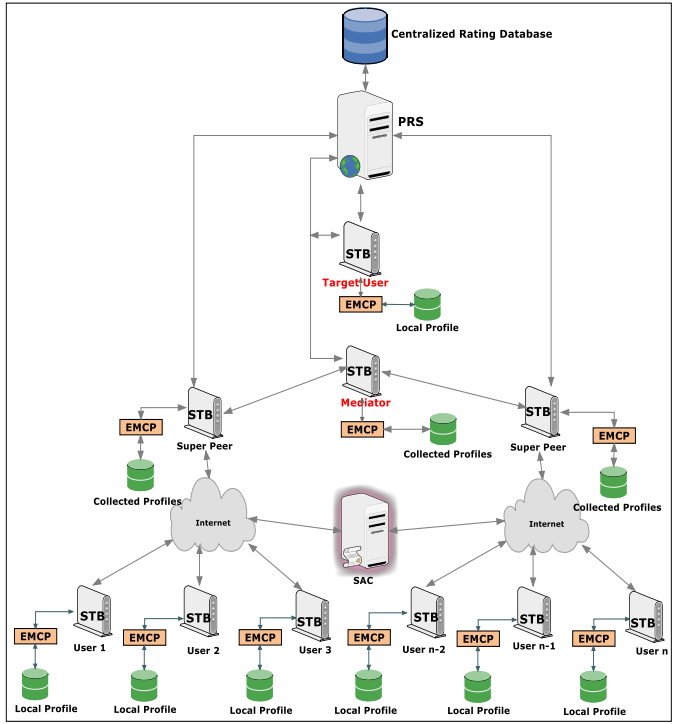

**Figure 1**: IPTV Network with Third Party Private Recommender Service

We don't assume the PRS to be completely malicious. This is a realistic assumption because PRS needs to accomplish some business goals and increase its revenues. Intuitively, the system privacy is high if PRS is not able to reconstruct the real ratings for users based on the information available to it. Figure (1) shows the architecture of our approach. Our Solution relies on hierarchical topology proposed in [22]; where participants are organized into peer-groups managed by super-peers. Electing super-peers is based on negotiation between participants and security authority centre. Security authority centre (SAC) is a trusted third party responsible for generating certificates for both the target-user and mediator, and managing these certificates. In addition, SAC is responsible for making assessment on those super-peers according to participants' reports and periodically update the reputation of these super-peers. The reputation mechanisms are employed to elect suitable super-peers based on estimating values for user-satisfaction, trust level, processing capabilities and available bandwidth, detailed and complex reputation mechanisms can be found in [23]. When a problem with specific super-peer occurs during the recommendation process, a participant can report it to SAC. After investigation, the assessment of the super-peer will be degraded. This will limit the chance for electing it as a super-peer in the future. On the other hand, successful recommendations processes will help to upgrade the super-peer reputation. IPTV provider can give certain benefits (like free content, prizes... etc) for those participants who have sustainable success rate as super-peers.

Our solution depends upon the Set top box (STB) device at user side. STB is an electronic appliance that connects to both the network and the home television. With the advancement of data storage technology each STB is equipped with a mass storage, e.g. Cisco STB. *EMCP* components are hosted on STB; Moreover STB storage stores the user rating profile. On the other hand, PRS maintains a centralized rating database that is used to provide recommendations if the number of participants below a certain threshold. PRS is the third-party entity recruited by the IPTV network provider to operate recommendations by consolidating the information received from multiple sources. We alleviate the user's identity problems by using anonymous pseudonyms identities for participants.

### IV. PROPOSED SOLUTION

In the beginning, we want to introduce the notions of privacy and trust within our framework, we need to justify what we mean by privacy and trust first. To define privacy and trust in our terms, we first approach the notion of privacy in following terms: "A target user who wants recommendations in a network of users, does not has to reveal raw ratings in his/her profile during the recommendations process and other users in the network can't learn any ratings in his/her raw profile". While in the context of this paper, trust is interpreted as "a user's expectation of another user's competency in providing ratings to reduce its uncertainty in predicating new items' ratings [24]". In our framework, the notion of privacy surrounding the disclosure of users' rating profiles and the protection of trust computation between different users together are the backbone of our solution. We apply multi-level obfuscation mechanism that produces different copy of participant's rating profile based on the different trust levels for target users. The trust computation is done locally over obfuscated users' rating profiles, and then recommendations are served using secure multi-party computation protocol. Utilizing Trust heuristic as an input for both group formation and multi-level obfuscation has been of great importance to mitigate some of malicious insider attacks such as infesting the trust computation results. As a future work, we plan to investigate miscellaneous insider attacks and strengthen our framework against them.

In the next sub-sections, we will present our proposed middleware for protecting the privacy of users' rating profiles

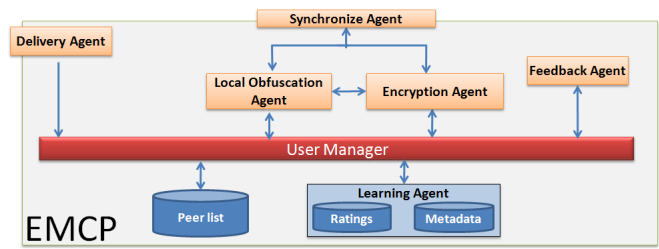

**Figure 2**: *EMCP* Components

Figure (2) demonstrates *EMCP* components running in the user's STB. *EMCP* consists of different co-operative agents. A Learning agent captures user ratings about items explicitly or implicitly to build a rating database and meta-data database. The local obfuscation agent implements multi-level obfuscation mechanism to achieve user privacy while sharing his/her rating profile with super-peers or PRS. The encryption agent is only invoked if the user is acting as a super-peer in the recommendation process; it executes *SR* protocol on the collected rating profiles. These mechanisms act as wrappers that obfuscate items' ratings before they are shared with any external entity. Since the database is dynamic in nature, the local obfuscation agent desensitizes the updated data periodically, and then synchronize agent forwards it to the PRS upon owner permit. So that recommendations to be made on the most recent ratings.

The recommendation process in our solution operates as follows:

1. The learning agent collects user's ratings about different items which represent his profile. The local profile is stored in two databases, the first one is the rating database that contains (item_id, rating) and the second is the meta-data database that contains feature vector for each item [25] (item_id, feature1, feature2, feature3). The feature vector can include genres, directors, actors and so on. Both implicit and explicit ways for information collection [26] are used to construct these two databases and maintain them. Clustering user's profile [27] is an extra step done by learning agent to reduce response time for different recommendation requests.

2. As stated in [18], the target user broadcasts a message to other users in the IPTV network to request recommendations for specific genre or category of items. Then he uses local obfuscation agent to sanitize the rest of items' ratings in local profile. In order to hide items identifiers and meta-data from other participants, The manger agent uses locality-sensitive hashing (LSH) [28] to hash these values. One interesting property for LSH is that similar items will be hashed to the same value with high probability. Super-peers and PRS are still be able to perform computation on the hashed items using appropriate distance metrics like hamming distance or dice coefficient. Finally, the target user dispatches these obfuscated items' ratings along with their associated hashed values to the Individual users who decided to participate in the recommendation process. These ratings are used in the computation of trust level at participant side.

3. Each group of participant negotiates with SAC to select a peer with highest reputation as a "super-peer" which will act as a communication gateway between the target user and participants in its underlying peer group.

4. The preparation phase of *SR* protocols starts such that the target user and super-peers need to generate a cryptographic key independently. One of the super-peers will act as a mediator that has to generate an encryption key *Mpk* then broadcast it to the target user and all super-peers. The target user initiates the process by sending his encryption key $\varepsilon_{Mpk}(\varepsilon_{Tpk})$ to the mediator; the mediator in turn decrypts the received value to obtain *Tpk*. Both parties exchange their encryption keys while their decryption keys are kept private and are not shared with the other participants. Next the mediator broadcasts *Tpk* to all super-peers. At the end of this phase, all super-peers hold the two encryption keys that will be used to doubly encrypt the aggregated ratings in the next phase. The target-user doubly encrypts his/her mean rating over rated items with his/her encryption key and the mediator key. Then, he/she submits this encrypted value to PRS.

5. The process of calculating interpersonal trust with target user is done in decentralized fashion using entropy definition proposed in [24] at each participant side. The entropy value becomes lower as users' ratings are more consistent, which is similar to the definition of trust stated before.

$\forall_{j=1}^{n} T(u_a, u_{b_j})$ is the estimated trust between the target user $u_a$ and participant $u_{b_j}$, it is computed privately using following steps:

i. Each participant $\forall_{j=1}^{n} u_{b_j}$ determines a subset of his/her items' ratings that will be required for recommendation process. Then participant utilizes shared items rated by both of them $u_a, u_{b_j}$ for the trust computation. Determining shared rated items is done by matching the received items' hash values from target user $u_a$ with his/her local items' hash values.

ii. Participant $u_{b_j}$ computes trust level using equation
$$T(u_a, u_{b_j}) = \frac{Entropy(u_a) - Entropy(u_a|u_{b_j})}{Entropy(u_a)} \quad (1)$$
$$= \frac{\left(1 - \frac{\log N}{\log ZN}\right) + \frac{1}{N \log ZN}\left(\sum_{i=1}^{Z}\sum_{j=1}^{Z} n_{ij} \log n_{ij} - \sum_{i=1}^{Z} n_i \log n_i\right)}{1 - \frac{1}{N \log ZN} \sum_{i=1}^{Z} n_i \log n_i}$$

Equation (1) is an adapted formalization of trust proposed in [24] where *Z* denotes the number of states of rated values, *N* is the total number of rating times. For example if *Z=6* and *N=20* when 20 ratings are made with 1 to 6 integer valued scores. Employing entropy to select trustworthy neighbours attains an improvement in group formation and rating predication. Enhancement in rating predication stemmed from trust propagation, so if $u_{b_j=x}$ selected as a trusty user and he/she doesn't have ratings for the item to be predicted, a trusty user $u_{b_j=y}$ of user $u_{b_j=x}$ can also be used for the predication.

iii. Each participant $\forall_{j=1}^{n} u_{b_j}$ sends his/her calculated trust value to super-peer. The Estimated trust values are forwarded to both super-peers and PRS.

6. Each participant $\forall_{j=1}^{n} u_{b_j}$ uses their local obfuscation agent to perform a multi-level obfuscation on items' ratings that are required in the recommendation process. Moreover the manager agent hashes their identifiers and meta-data using LSH. The level of obfuscation is determined using the trust level with the target user, and then participants submit their locally obfuscated profiles to the super-peer of their group. Secure routing protocols [29] can be utilized to hide the network identities of group members when submitting their locally obfuscated profiles to the super-peers.

7. Upon receiving the obfuscated profiles from the participants, each super-peer filters the received profiles based on the trust level of their owners such that $T(u_a, u_{b_j}) > \theta$ where $\theta$ is a minimum trust threshold value defined by the target user. Then, each super-peer collects participants' pseudonyms and aggregates group profiles such that all the <hashed value, rating> elements belonging to similar items clustered together. This allows computing items popularity curve at each super-peer. The super-peer can seamlessly interact with the PRS by posing as an end-user has group profile as his own profile. each super-peer $\forall_{x=1}^{k} SP_x$ calculates the following intermediate values for each user in the *N*-neighbourhood of target user $\forall_{j=1}^{n} u_{b_j} \in \text{Neighbor}(u_a)$,
Then $\forall\, q = 1 \ldots T \quad \widetilde{r_{u_{b_j},q}} = r_{u_{b_j},q} - \overline{r_q}$

$$\widehat{r_{q,u_{b_j}}}x = \frac{T(u_a,u_{b_j})*\widehat{r_{u_{b_j},q}}}{T(u_a,u_{b_j})} \quad (2)$$

Where $r_{u_{b_j},q}$ is the rating value of participant $u_{b_j}$ for item $q$. $\overline{r_q}$ is the average rating for item $q$ in each items' cluster. Next, each super-peer performs a doubly encryption on the intermediate ratings $\widehat{r_{q,u_{b_j}}}x$ of each participant using the encryption key of the target user $Tpk$ and mediator $Mpk$ (encryption phase). Finally, the super-peer submits these ratings along with their associated hashed values to PRS, which in turn collects them to produce final referrals.

8. Upon receiving the doubly encrypted ratings $\forall_{x=1}^{k} \forall_{j=1}^{n} \varepsilon_{Mpk}\left(\varepsilon_{Tpk}\left(\widehat{r_{q,u_{b_j}}x}\right)\right)$ from all super-peers, PRS stores them along with their participants' pseudonyms and hashed values in the centralized rating database. The recommendation phase is performed using additive homomorphic property of the paillier encryption as the required computations are addition. Thus, PRS executes an additive operation on the doubly encrypted rating profiles without decrypting them so the private data of multiple super-peers can be preserved during the computation. Calculating predicted rating for referrals done as shown in equation (3):

$$p_{u_a,q} = \varepsilon_{Mpk}\left(\varepsilon_{Tpk}(\overline{r_{u_a}})\right) * \left(\prod_{j=1}^{n} \varepsilon_{Mpk}\left(\varepsilon_{Tpk}\left(\widehat{r_{q,u_{b_j}}x}\right)\right)\right)$$
$$= \varepsilon_{Mpk}\left(\varepsilon_{Tpk}\left(\overline{r_{u_a}} + \left(\sum_{j=1}^{n} \widehat{r_{q,u_{b_j}}x}\right)\right)\right) \quad (3)$$

Notice that the result will be equal to the weighted sum of participants' rating plus the average rating for target user $r_{u_a}$. PRS uses the reblinding property of paillier encryption to prevent the mediator from getting knowledge of $p_{u_a,q}$ values before sending them back to the target user by trying a few possible values.

9. PRS forwards the doubly encrypted referrals list along with their predicated ratings to the mediator that in turn decrypts it and forwards the output to the target user. The target-user in turn decrypts the list to get the final output because he/she holds the final decryption key. Optionally, the target-user publishes the final list to other participants in the recommendation process. Finally, each participant report scores about the elected super-peer of his group and target-user to SAC, which helps to determine reputation of each entity involved in referrals generations.

## V. PROPOSED TWO STAGE CONCEALMENT PROCESS

In the next subsections, we present a two stage concealment process used in *EMCP* to disguise the user rating profile in a way that secure user's ratings in the untrusted PRS with minimum loss of accuracy. In our framework, each user has two datasets representing his/her profile. Local profile: it represents the actual ratings of the user for different items; it is stored on his STB. Each user disguises this local profile before sending it to super-peer. A centralized profile: this is the output of the two-stage concealment process where the user gets recommendation directly from the PRS based on the previously collected profiles. We perform experiments on real datasets to illustrate the applicability of our mechanisms and the privacy and accuracy levels achieved using them.

### A. Cryptigrpahy Tools

We employ homomorphic encryption scheme to preserve the privacy of ratings collected by super-peers. Moreover, homomorphic encryption possesses specific properties that permit computation of linear combinations of encrypted data without need for prior decryption. Formally, an encryption schema $\varepsilon_{pk}(.)$ denotes the encryption function with encryption key $pk$ and $D_{sk}(.)$ denoted the decryption function with decryption key $sk$. Additive homomorphic cryptosystem possesses the following properties:

1. Given the encryption of plaintexts $m_1$ and $m_2$, $\varepsilon_{pk}(m_1)$ and $\varepsilon_{pk}(m_2)$. The sum $m_1 + m_2$ can be directly computed as $\varepsilon_{pk}(m_1 + m_2) = \varepsilon_{pk}(m_1) * \varepsilon_{pk}(m_2)$.

2. Given a constant $k$ and the encryption of $m_1$, $\varepsilon_{pk}(m_1)$. The multiplication of $k$ with the plaintext $m_1$ can be directly computed as $\varepsilon_{pk}(k.m_1) = \varepsilon_{pk}(m_1)^k$.

In designing SR protocol we used Paillier cryptosystem as it provides a strong security due to the use of randomized encryption so an adversary cannot even see whether two encryptions correspond to the same text. we will briefly state the Paillier cryptosystem, a more detailed description can be found in [30].

**Key Generation**
In this step, two large primes $p$ and $q$ are chosen randomly where $p < q$ and $p \nmid q - 1$. The encryption key $pk$ is set to $N$ where $N = p.q$, and the decryption key $sk$ is set to $(\lambda, N)$ where $\lambda = lCM(p - 1, q - 1)$.

**Encryption**
Given $n$, $g \in \mathbb{Z}_{N^2}^*$ is a generator whose order divides N, plaintext $m$ and a random number $r \in [1, \ldots, N - 1]$. The encryption of the message $m \in \mathbb{Z}_n : \varepsilon_{pk}(m) = g^m . r^N \bmod N^2$. For any encrypted message, a different encryption can be computed by multiplying it with a random blinding factor $r^N$.

**Decryption**
Given $N$, the cipher-text $c = \varepsilon_{pk}(m)$, the decryption is as follows $m = \frac{(c^\lambda \bmod N^2) - 1}{N} \lambda^{-1}$ where $\lambda^{-1}$ is the inverse of $\lambda$ in module $N$.

### B. Local Obfuscation using Clustering Transformation Algorithm (CTA)

We propose a novel algorithm for obfuscating the user rating profile before sending it to super-peer. This algorithm called *CTA*, which has been designed especially for the sparse data problem we have here, Moreover the algorithm permits multi-level obfuscation based on trust level. We noted that, the available anonymisation algorithms increase data distortion as they perform single obfuscation level for all participants and release one obfuscated copy for all of them, as result inaccurate recommendation model could constructed. Maintaining utility and privacy for profiles seems to be contradictory goals to attain. *CTA* partitions the user profile into smaller clusters and then pre-process each cluster such that the distances inside the same cluster will maintained in its obfuscated version. We use local learning analysis (*LLA*) clustering method proposed in [31] to partition the dataset. After complete the partitioning, we embed each cluster into a random dimension space so the sensitive ratings will be protected. Then the resulting cluster will be rotated randomly. In such a way, *CTA* obfuscates the data inside user profile while preserving the distances between the data points to provide accurate results when performing recommendations. The output of our obfuscation algorithm should satisfy two requirements:

- Reconstructing the original profile from the obfuscated profile should be difficult, in order to preserve privacy.
- Preserving the distances of the data to achieve accurate results for the recommendations.

Our algorithm consists of the following steps:

1. The user ratings is stored in STB as dataset $D$ consists of $c$ rows, where each row is a sequence of $X$ attributes where $X = x_1\ x_2\ x_3 \ldots \ldots x_n$.
2. The dataset $D$ is portioned vertically into $D_1\ D_2\ D_2 \ldots \ldots D_m$ subsets of length $L$, if $n/L$ is not perfectly divisible then *CTA* randomly selects attributes already assigned to any subset and joins them to the attributes of the incomplete subsets.
3. Cluster each subset $\forall_{j=1}^{m} D_j$ Using *LLA* algorithm, which resulting in $K$ clusters $D_j = C_{j1}, C_{j2}, C_{j3} \ldots C_{jk}$ for each subset. Note that *LLA* uses Gaussian Influence function as the similarity measure. Influence function between two data point $x_i$ and $x_j$ is given by

$$f_{Gauss}^{x_i}(x_j) = e^{-\frac{d(x_i,x_j)^2}{2\sigma^2}} \quad (4)$$

And the field function for a candidate core-point selection, which is given by:

$$f_{Gauss}^{D}(x_j) = \sum_{s=1}^{k} e^{-\frac{d(x_j,x_{is})^2}{2\sigma^2}} \quad (5)$$

So every point in the original dataset $D$ falls exactly in one cluster. The aim of this step is to increase the privacy level of the transformation process and make reconstruction attacks difficult.

4. *CTA* generates two sets for each cluster $\forall_{i=1}^{k} C_{ji}$ in the subset $D_j$ these are $H_{ji}$ and $O_{ji}$. Where $H_{ji}$ is the set of points with highest values for field function and $O_{ji}$ is the rest of points in $C_{ji}$. For each point $x_{1i} \in H_{ji}$ construct a weighted graph $\Gamma_i$ that contains its *k*-nearest neighbours in $O_{ji}$, each edge $e \in \Gamma_i$ has a weight equals to $f_{Gauss}^{b_{1i}}(x_{1i})$.

5. Estimate the geodesic distances by Computing the shortest distance between each two points in graph $\Gamma_i$ using Dijkstra or Floyd algorithm and then build a distance matrix $D_{\Gamma_i} = \{f_{Gauss}^{b_i}(x_i)\}$.

6. Based on $D_{\Gamma_i}$, we find a *d-dim* embedding space $C'_{ji}$ using classical *MDS* [32] as follows

    - Calculate the matrix of squared distances $S = D_{\Gamma_i}^2$ and the centering matrix $H = 1 - 1/N\ ee^T$
    - The characteristic vectors are chosen to minimize $E = \|\tau(D_{\Gamma_i}) - \tau(D_d)\|_{L^2}$, where $\tau(D_d)$ is the distance matrix for the *d-dim* embedding space, and $\tau$ converts distances to inner products $\tau = -HSH/2$.
    - Trust level values are divided to a number of intervals defined by the user, associated with each interval a *d-dim* value. *CTA* chooses a value for *d-dim* according to the trust level associated with the target user.

7. For each cluster $\forall_{j=1}^{m} \forall_{i=1}^{k} C'_{ji}$, *CTA* randomly select two attributes $x_a$ and $x_b$ to perform rotation perturbation on selected attributes $R(x_a, x_b)$ using transformation matrix $M_j^\theta$ setup by the user for each cluster using range of angles defined in advance by the user.

8. Repeating steps 4-7 for all clusters in $\forall_{j=1}^{m} D_j$ to get the obfuscated portion $D'_j$. Finally, the obfuscated dataset is obtained by $D' = \cup_{j=1}^{n} D'_j$.

*C. Secure Recommendation Protocol (SR)*

We proposed a protocol that enables PRS to calculate predicted ratings from the doubly encrypted profiles. We called this protocol secure recommendation protocol (*SR*). *SR* consists of three phases: preparation phase, encryption phase and recommendation phase as stated in section IV. Employing a multi-level obfuscation mechanism on local profiles based on the estimated trust level with target-user ensures participants' privacy as the local profile only leaves participant's device desensitised. Obfuscating local profiles permits super-peers to perform intermediate computation on their obfuscated shares of ratings without a need for extra SMC protocols or large encryption keys. Moreover, obfuscation requires lees computational and communication resources than that required for encryption. The final recommendation phase is carried out in form of secure addition of encrypted ratings profiles received from super-peers.

1. Preparation phase: generation of cryptographic key pair by mediator and target user
    For mediator
        Generate encryption key $Mpk$
        Broadcast $Mpk$ to all super-peers and target user
        Broadcast $Tpk$ to all super-peers
    End for
    For Target-user
        Generate encryption key $Tpk$
        Send $\varepsilon_{Mpk}(\varepsilon_{Tpk})$ to mediator
    End for

2. Encryption phase: Each super-peer encrypts his/her aggregated ratings profiles with both encryption keys
    For each participant $\forall_{j=1}^{n} u_{b_j}$ do
        Extracts his/her rating for requested item $r_{u_{b_j},q}$
        Calculates the trust level with target user $T(u_a, u_{b_j})$
        Perform multi-level obfuscation on $\forall_{T}^{1} r_{u_{b_j},q}$
        Send $r_{u_{b_j},q}$, $T(u_a, u_{b_j})$ to the super-peer in his group
    End for
    For each super-peer $\forall_{x=1}^{k} SP_x$ do
        Calculates $\widetilde{r_{q,u_{b_j}}^{x}} = \frac{T(u_a, u_{b_j}) * \widetilde{r_{u_{b_j},q}}}{T(u_a, u_{b_j})}$
        Send $\widetilde{r_{q,u_{b_j}}^{x'}} = \varepsilon_{MpK}(\varepsilon_{TpK}(\widetilde{r_{q,u_{b_j}}^{x}}))$ to PRS
    End for

3. Recommendation phase: PRS generates referrals by accumulating the received shares in order to attain a predicated rating for each referred item. detailed steps can be stated as follows:

    PRS receives $\widetilde{r_{1,u_{b_1}}^{1'}}, \widetilde{r_{1,u_{b_1}}^{2'}}, \ldots, \widetilde{r_{2u_{b_1}}^{1'}}, \widetilde{r_{2u_{b_2}}^{2'}}, \ldots, \widetilde{r_{Tu_{b_j}}^{k'}}$ such that $\widetilde{r_{qu_{b_j}}^{k'}}$ is the doubly encrypted rating for item $q \in \{1, \ldots T\}$ by user $u_{b_j} (\forall_{j=1}^{n} u_{b_j} | T(u_a, u_{b_j}) > \theta)$ from super-peer $SP_x (\forall_{x=1}^{k} SP_x)$

For each item $q = 1$ to T do
PRS Calculates $\forall_{x=1}^{k} p_{u_a,q} = \varepsilon_{Mpk}\left(\varepsilon_{Tpk}(\widehat{r_{u_a}})\right) * \left(\prod_{j=1}^{n} \varepsilon_{Mpk}\left(\varepsilon_{Tpk}\left(\widehat{r_{q,u_{b_j}}^x}\right)\right)\right) \forall_{j=1}^{n} u_{b_j}$
End for

PRS send the list of items and their predicated ratings $\{(item_1, p_{u_a,1}), (item_2, p_{u_a,2}), \ldots (item_T, p_{u_a,T})\}$ to the mediator. Next, the mediator will decrypt the received list and sends it to the target user. The target-user is able to find the final output because it holds the final decryption key. The target user filters out the received list on his device by removing items with a low predicated rating and items already consumed items.

## VI. Proof of Security and Correctness

The proof of security for *SR* protocol depends on how much information is leaked during the execution of recommendation phase. At the same time, our *SR* protocol should output accurate results.

**Theorem 1:** additive operation performed by PRS in *SR* protocol is correct and accurate without the need of decryption keys.

**Proof:** based on the first property of Additive homomorphic cryptosystem, we can derive that additive operations for doubly encrypted data are correct as follow: Given the encryptions $\varepsilon_{pk1}(m_1) = a$ and $\varepsilon_{pk1}(m_2) = b$ where $\forall m_1, m_2 \in \mathbb{Z}_n$, given encryption key $pk2$
$\varepsilon_{pk2}\left(\varepsilon_{pk1}(m_1)\right) \cdot \varepsilon_{pk2}\left(\varepsilon_{pk1}(m_2)\right) \mod N^2 = \varepsilon_{pk2}(a) \cdot \varepsilon_{pk2}(b) = (g^a r_1^N) \cdot (g^b r_2^N) \mod N^2 = g^{a+b}(r_1 r_2)^N \mod N^2 = \varepsilon_{pk2}(\varepsilon_{pk1}(m_1 + m_2)) \mod N^2$
Based on that, the PRS does not require any decryption key in order to aggregate all encrypted data.

**Theorem 2:** *SR* protocol computes predicated ratings for each referred item based on similar users' ratings without revealing extra information to any party.

**Proof:** Since each participant obfuscates items' ratings and hashes their meta-data prior to submitting them to super-peer. Moreover, each super-peer encrypts the aggregated profiles with the encryption keys of the target user and mediator. No single party is able to decrypt the encrypted profiles. In our two stage concealment process, the super-peer aggregates all obfuscated profiles then performs intermediate-computations on the obfuscated ratings for each item without having to know their real ratings. No party can see extra information during the execution of *SR* protocol except the target user at the end of protocol. As for participants, they only participate in the recommendation process without knowing other participants' identity. As neither all participants have the same super-peer nor do have direct communication with each other. The local profile is secured and can only be viewed by its owner before applying multi-level obfuscation mechanism. In addition, employing reputation techniques to select super-peers with high success rate in previous recommendations processes ensures selection of reliable peers that will perform required phases. PRS can't see the received profiles as none of the decryption keys are known. Furthermore, the decryption process requires both decryption keys stored at the target-user and mediator. After PRS generates final referrals list, PRS submits it to the mediator which in turn perform the first decryption process. After the first decryption, the mediator is not able to see the final result because the final decryption key is held by the target user.

**Theorem 3:** Assuming that all parties follow the protocol, *SR* protocol can correctly computes the predicated rating for each referred item.

**Proof:** When each super-peer encrypts his aggregated profiles with both encryption keys $\varepsilon_{MpK}\left(\varepsilon_{TpK}\left(\widehat{r_{q,u_{b_j}}^x}\right)\right)$. PRS performs additive operation on doubly encrypted profiles based on Paillier's homomorphic cryptosystem as follows:

$p_{u_a,q} = \varepsilon_{Mpk}\left(\varepsilon_{Tpk}(\widehat{r_{u_a}})\right)$
$\quad + \left(\varepsilon_{Mpk}\left(\varepsilon_{Tpk}(\widehat{r_{q,1}})\right) + \varepsilon_{Mpk}\left(\varepsilon_{Tpk}(\widehat{r_{q,2}})\right)\right.$
$\quad + \varepsilon_{Mpk}\left(\varepsilon_{Tpk}(\widehat{r_{q,3}})\right) + \cdots$
$\quad \left. + \varepsilon_{Mpk}\left(\varepsilon_{Tpk}(\widehat{r_{q,u_{b_j}}})\right)\right)$

$p_{u_a,q} = \varepsilon_{Mpk}\left(\varepsilon_{Tpk}\left(\widehat{r_{u_a}} + \left(\sum_{j=1}^{n} \widehat{r_{q,u_{b_j}}^x}\right)\right)\right)$

$p_{u_a,q} = \varepsilon_{Mpk}\left(\varepsilon_{Tpk}(\widehat{r_{u_a}})\right) * \left(\prod_{j=1}^{n} \varepsilon_{Mpk}\left(\varepsilon_{Tpk}\left(\widehat{r_{q,u_{b_j}}^x}\right)\right)\right)$ (6)

After the first decryption by mediator, we have
$p_{u_a,q} = \varepsilon_{Tpk}(\widehat{r_{u_a}}) * \left(\prod_{j=1}^{n} \varepsilon_{Tpk}\left(\widehat{r_{q,u_{b_j}}^x}\right)\right)$ (7)

When the target-user performs the last decryption, he will obtain the final predicated rating as in equation (8)
$p_{u_a,q} = \widehat{r_{u_a}} * \left(\prod_{j=1}^{n} \widehat{r_{q,u_{b_j}}^x}\right)$ (8)

So the result from *SR* protocol is correct.

## VII. Experiments

In this section, we describe the implementation of our proposed solution. The experiments run in Intel® Core 2 Duo™ 2.4 GHz processor with 2 GB Ram. We used MySQL as a data storage. The proposed two stage concealment process implemented in C++. We used message passing interface (MPI) for the distributed memory implementation of *SR* protocol to mimic a distributed reliable network of peers. The experiments presented here were conducted using the Jester dataset provided by Goldberg from UC Berkley [33]. The dataset contains 4.1 million ratings on jokes using a real value between (-10 and +10) of 100 jokes from 73.412 users. The data in our experiments consists of ratings for 36 or more items by 23.500 users. We evaluated the proposed solution from two different aspects: privacy achieved and accuracy of results. We used the mean absolute error (MAE) metric proposed in [34]. MAE is one of most famous metrics for recommendation quality. We can define it as following: Given a user predicated ratings set $p = \{p_1, p_2, p_3 \ldots p_N\}$ and the corresponding real ratings set $r = \{r_1, r_2, r_3 \ldots r_N\}$, MAE is:

$$MAE = \sum_{i=1}^{N} |p_i - r_i| / N \quad (9)$$

MAE measures the predication verity between the predicated ratings and the real ratings, so smaller MAE means better recommendations provided by PRS. To measure the privacy or distortion level achieved using our mechanism, we used variation of information metric VI [35] to estimate data error. VI is:

$$VI = H(p) + H(r) - 2I(p,r) \quad (10)$$

Here $H(p)$ is entropy of $p, r$ and $I(p,r)$ is mutual information between $p$ and $r$. The higher VI means the larger distortion between the obfuscated and original dataset, which means higher privacy level.

The experiments involve dividing the data set into a training set and testing set. The training set is obfuscated then used as a database for PRS. Each rating record in the testing set is divided into rated items $t_i$ and unrated items $r_i$. The set $t$ is presented to the PRS for making predication $p_i$ for the unrated items $r_i$. For the representation process of trust calculation, we add the default value 0 for the items not rated. In our dataset, the first column of every raw store how many items are rated by the user, which is necessary for trust estimation process. We fix the number of super-peers to be 3, as described earlier they will be responsible on aggregating the data of 23.496 participants. We assume the trust level for all participants above minimum trust threshold $\theta$ that is required for the inclusion in predication process. The recommendation process can be initiated by any user that will be act as target-user for the referrals list. The trust level between participants and target-user is calculated locally on their STB devices.

We used our *SR* protocol to predicate referred items' ratings based on the weighted ratings for each participant. First we want to measure the encryption performance in *SR*, so we collect all 23.496 records on one super-peer and doubly encrypt the aggregated data with different encryption key length of 128, 256, 512, 1024 and 2048 referring to figure (3). With 128 bits, the time elapsed by encryption of 23.496 records is about 3.120 ms and 4.230 for 256 bits, 5.814 for 512 bits 8.164 ms for 1024 bits and 12.241ms for 2048 bits respectively. As we can see, this result presents an exponential cost of time while doubling the encryption key length.

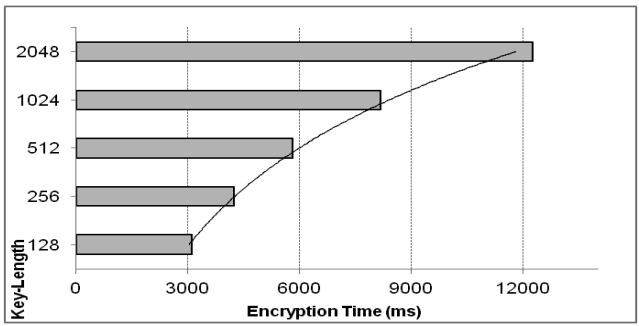

**Figure 3**: Encryption Time versus Key Length.

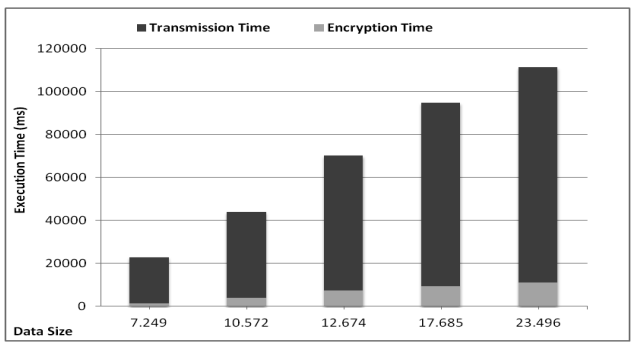

**Figure 4**: Execution Time for Different Data Size

In the second experiment, we vary minimum trust threshold to attain different number of participants' records in the recommendation process, then we run SR protocol on these aggregated records in size of 7.249, 10.572, 12.674, 17.685, and 23.496. As shown in figure (4), we can see that both encryption time and transmission time grow linearly while we enlarge the data size. Moreover, we can see that, the execution time is directly proportional with data size and transmission time is dominating this increase in total execution time cost.

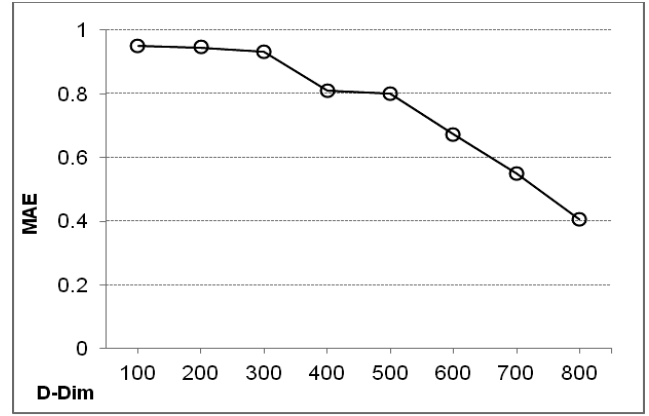

**Figure 5:** Accuracy of recommendations for obfucated dataset using *CTA*

To evaluate the accuracy of *CTA* algorithm with respect to different number of dimensions in rating profile, we control *d-dim* parameters of *CTA* to vary number of dimensions during the evaluation. Figure (5) shows the performance of recommendations of locally obfuscated data, as shown the accuracy of recommendations based on obfuscated data is low when the dimension is low. But at a certain number of dimensions (500), the accuracy of recommendations of obfuscated data is nearly equal to the accuracy obtained using original data.

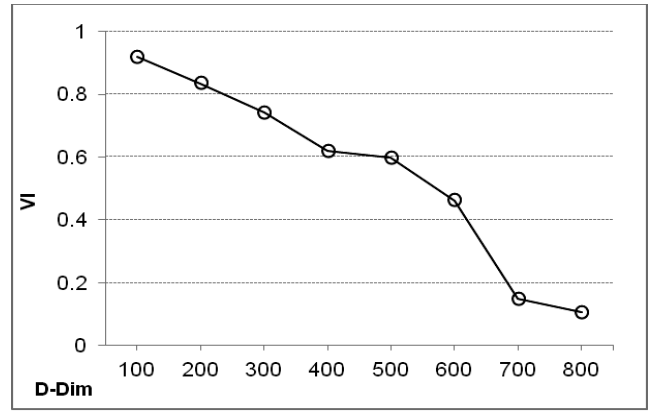

**Figure 6:** Privacy levels for the obfucated dataset using *CTA*

In the second experiment performed on *CTA* algorithm, we examine the effect of *d-dim* on VI values. As shown in figure (6), VI values decrease with respect to the increase in *d-dim* values for rating profile. *d-dim* is the key element for privacy level where smaller *d-dim* value, the higher VI values (privacy level) of *CTA*. However, clearly the highest privacy is at *d-dim*=100. There is a noticeable drop of VI values when we change *d-dim* from 300 to 600. *d-dim* value 400 is considered as a critical point for the privacy. Note that rotation transformation adds extra privacy layer to the data and in the same time maintains the distance between data points to enable PRS to build accurate recommendation models.

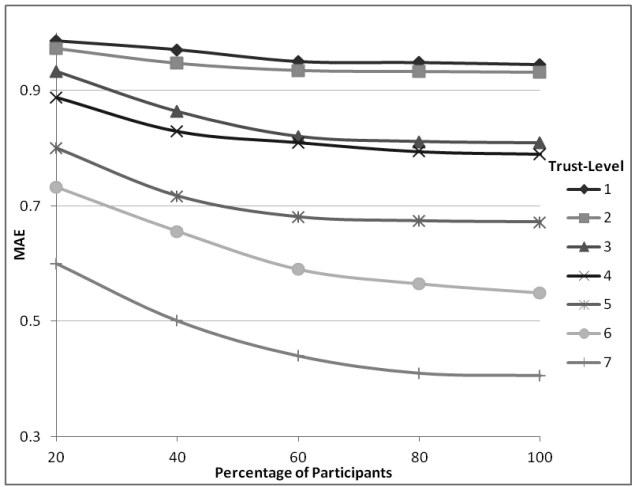

**Figure 7**: MAE Values With Different Percetage of Participants.

In the final experiment, we want to measure the impact of varying trust level and number of participants on the accuracy of the recommendations. We simulate a general case where the number of users was fixed to be 23.496. Then we assign different number of participants to a certain recommendation request, and gradually increased the percentage of users who joined the request from 10% to 100% of them. We fixed trust level for the target user with participants for each simulation (Trust level has an impact on *CTA* parameters) then we measured MAE for the results. As shown in figure (7), the MAE value occurs at approximately 60% of the participants with high trust levels are close to the MAE value for all users. Our conclusion is that, a low percentage of participants employing multi-level obfuscation with high trust level can attain MAE value close to the original MAE value obtained from running the recommendation process for all users. As a result the target user does not need to broadcast the request to the full IPTV network to attain accurate results but he can employ multicast for trusty users stored in his peer list to reduce the load in the network traffic. To illustrate the decrement of MAE values for predications based on diverse percentages of participants and trust levels, we calculated and plot figure (7). This verifies our conclusion that MAE approximately converges to the MAE which obtained using the whole users in our case.

## VIII. CONCLUSION AND FUTURE WOK

In this paper, we presented our ongoing work on building an enhanced middleware for collaborative privacy in IPTV recommender services. We gave a brief overview of *EMCP* architecture, the recommendations process with application to IPTV. We presented a novel two stage concealment process that provides to participants complete privacy control over their ratings profiles. The concealment process utilizes hierarchical topology, where participants are organized in groups, from which super-peers are elected based on their reputation. Super-peers aggregate the results obtained from underlying participants and then encapsulate intermediate values computed on these aggregated data and then send them to PRS. Multi-level obfuscation mechanism is used in the course of participant data collection, while *SR* protocol used to protect the privacy of collaborative filtering by distributing the participants' data between multiple super-peers and exchange only a subset of aggregated ratings which is useful for the recommendations. We test the performance of the proposed mechanisms on real dataset. We evaluated how the overall accuracy of the recommendations depends on number of participants and trust level. The experimental and analysis results showed that privacy increases under proposed middleware without hampering the accuracy of recommendations. In particular mean absolute error can be reduced with proper tuning of the multi-level obfuscation parameters for large number of participants. Moreover, utilizing trust level for multi-level obfuscation is an optimization to maintain the utility of rating profiles. Thus adding the proposed middleware does not severely affect the accuracy of recommendations based on collaborative filtering techniques.

We realized that there are many challenges in building a privacy enhanced middleware for recommender service. As a result we focused in middleware for collaborative privacy scenario. A future research agenda will include utilizing game theory to better formulate users groups, sequential profile release and its impact on privacy. Reducing transmission time and the load in the network traffic by adding secure filtering phase to *SR* protocol that allow PRS to exclude items with low predicated rating from final referrals list. Strengthen our middleware against shilling attacks, extend our scheme to be directed towards multi-dimensional trust propagation and distributed collaborative filtering techniques in p2p environment. Moreover, we need to investigate weighted features vector methods and its impact in released ratings. Such that, the participants not only obfuscates his items' ratings based on the trust level of target-user, but also he can express specific items to be diversely obfuscated with each trust level. We need to perform extensive experiments in other real datasets from UCI repository and compare the performance with other techniques proposed in the literature. Finally we need to consider different data partitioning techniques as well as identify potential threats and add some protocols to ensure the privacy of the data against those threats.

ACKNOWLEDGMENT

This work has received support from the Higher Education Authority in Ireland under the PRTLI Cycle 4 programme, in the FutureComm project (Serving Society: Management of Future Communications Networks and Services).

REFERENCES

[1] K. Kawazoe, *et al.*, "Platform Application Technology Using the Next Generation Network," NTT 2007.

[2] Z. Huang, *et al.*, "Applying associative retrieval techniques to alleviate the sparsity problem in collaborative filtering," *ACM Trans. Inf. Syst.,* vol. 22, pp. 116-142, 2004.

[3] L. Ardissono, *et al.*, *Personalized Digital Television: Targeting Programs to Individual Viewers (Human-Computer Interaction Series, 6)*: Kluwer Academic Publishers, 2004.

[4] M. d. Gemmis, *et al.*, "Preference Learning in Recommender Systems," presented at the European Conference on Machine Learning and Principles and Practice of Knowledge Discovery in Databases (ECML/PKDD), Slovenia, 2009.

[5] F. McSherry and I. Mironov, "Differentially private recommender systems: building privacy into the net," presented at the Proceedings of the 15th ACM SIGKDD international conference on Knowledge discovery and data mining, Paris, France, 2009.

[6] A. Esma, "Experimental Demonstration of a Hybrid Privacy-Preserving Recommender System," 2008, pp. 161-170.

[7] J. Canny, "Collaborative filtering with privacy via factor analysis," presented at the Proceedings of the 25th annual international ACM SIGIR conference on Research and development in information retrieval, Tampere, Finland, 2002.

[8]  J. Canny, "Collaborative Filtering with Privacy," presented at the Proceedings of the 2002 IEEE Symposium on Security and Privacy, 2002.

[9]  H. Polat and W. Du, "Privacy-Preserving Collaborative Filtering Using Randomized Perturbation Techniques," presented at the Proceedings of the Third IEEE International Conference on Data Mining, 2003.

[10]  H. Polat and W. Du, "SVD-based collaborative filtering with privacy," presented at the Proceedings of the 2005 ACM symposium on Applied computing, Santa Fe, New Mexico, 2005.

[11]  Z. Huang, *et al.*, "Deriving private information from randomized data," presented at the Proceedings of the 2005 ACM SIGMOD international conference on Management of data, Baltimore, Maryland, 2005.

[12]  H. Kargupta, *et al.*, "On the Privacy Preserving Properties of Random Data Perturbation Techniques," presented at the Proceedings of the Third IEEE International Conference on Data Mining, 2003.

[13]  B. N. Miller, *et al.*, "PocketLens: Toward a personal recommender system," *ACM Trans. Inf. Syst.,* vol. 22, pp. 437-476, 2004.

[14]  C.-N. Ziegler, *et al.*, "Improving recommendation lists through topic diversification," presented at the Proceedings of the 14th international conference on World Wide Web, Chiba, Japan, 2005.

[15]  J. Golbeck and J. Hendler, "FilmTrust: movie recommendations using trust in web-based social networks," in *Consumer Communications and Networking Conference, 2006. CCNC 2006. 3rd IEEE*, 2006, pp. 282-286.

[16]  R. Parameswaran and D. M. Blough, "Privacy preserving data obfuscation for inherently clustered data," *Int. J. Inf. Comput. Secur.,* vol. 2, pp. 4-26, 2008.

[17]  A. M. Elmisery and D. Botvich, "An Agent Based Middleware for Privacy Aware Recommender Systems in IPTV Networks," in *Intelligent Decision Technologies*. vol. 10, J. Watada, *et al.*, Eds., ed: Springer Berlin Heidelberg, 2011, pp. 821-832.

[18]  A. Elmisery and D. Botvich, "Private Recommendation Service For IPTV System," in *12th IFIP/IEEE International Symposium on Integrated Network Management*, Dublin, Ireland, 2011.

[19]  A. Elmisery and D. Botvich, "Agent Based Middleware for Maintaining User Privacy in IPTV Recommender Services," in *3rd International ICST Conference on Security and Privacy in Mobile Information and Communication Systems*, Aalborg, Denmark, 2011.

[20]  A. Elmisery and D. Botvich, "Privacy Aware Obfuscation Middleware for Mobile Jukebox Recommender Services," in *The 11th IFIP Conference on e-Business, e-Service, e-Society*, Kaunas, Lithuania, 2011.

[21]  A. Elmisery and D. Botvich, "Privacy Aware Recommender Service for IPTV Networks," in *5th FTRA/IEEE International Conference on Multimedia and Ubiquitous Engineering*, Crete, Greece, 2011.

[22]  W. Nejdl, *et al.*, "Super-peer-based routing and clustering strategies for RDF-based peer-to-peer networks," presented at the Proceedings of the 12th international conference on World Wide Web, Budapest, Hungary, 2003.

[23]  J. Carbo, *et al.*, "Trust management through fuzzy reputation. Int," *Journal in Cooperative Information Systems,* vol. 12, p. 135—155, 2002.

[24]  H. D. Kim, "Applying Consistency-Based Trust Definition to Collaborative Filtering," *KSII TRANSACTIONS ON INTERNET AND INFORMATION SYSTEMS,* vol. 3, pp. 366-374, 2009.

[25]  A. Elmisery and D. Botvich, "Agent Based Middleware for Private Data Mashup in IPTV Recommender Services," in *16th IEEE International Workshop on Computer Aided Modeling, Analysis and Design of Communication Links and Networks*, Kyoto, Japan, 2011.

[26]  D. Kelly and J. Teevan, "Implicit feedback for inferring user preference: a bibliography," *SIGIR Forum,* vol. 37, pp. 18-28, 2003.

[27]  Y. Ye, *et al.*, "A comparative study of feature weighting methods for document co-clustering," *International Journal of Information Technology, Communications and Convergence,* vol. 1, pp. 206-220, 2010.

[28]  P. Indyk and R. Motwani, "Approximate nearest neighbors: towards removing the curse of dimensionality," presented at the Proceedings of the thirtieth annual ACM symposium on Theory of computing, Dallas, Texas, United States, 1998.

[29]  M. Imani, *et al.*, "Security enhanced routing protocol for ad hoc networks," *Journal of Convergence,* vol. 1, pp. 43-48, 2010.

[30]  P. Paillier, "Public-Key Cryptosystems Based on Composite Degree Residuosity Classes."

[31]  A. Elmisery and F. Huaiguo, "Privacy Preserving Distributed Learning Clustering Of HealthCare Data Using Cryptography Protocols," in *34th IEEE Annual International Computer Software and Applications Workshops*, Seoul, South Korea, 2010.

[32]  I. Borg and P. J. F. Groenen, *Modern Multidimensional Scaling: Theory and Applications (Springer Series in Statistics)*: Springer, 2005.

[33]  D. Gupta, *et al.*, "Jester 2.0 (poster abstract): evaluation of an new linear time collaborative filtering algorithm," presented at the Proceedings of the 22nd annual international ACM SIGIR conference on Research and development in information retrieval, Berkeley, California, United States, 1999.

[34]  J. L. Herlocker, *et al.*, "Evaluating collaborative filtering recommender systems," *ACM Trans. Inf. Syst.,* vol. 22, pp. 5-53, 2004.

[35]  C. Kingsford, "Information Theory Notes," 2009.